\documentclass[12pt]{article}%
\usepackage{ amsmath, graphics, rotating}% 

\begin{document}
\large
\begin{center}{\large\bf COULD ONLY FERMIONS BE ELEMENTARY?}
\end{center}
\vskip 1em \begin{center} {\large Felix M. Lev} \end{center}
\vskip 1em \begin{center} {\it Artwork Conversion Software Inc.,
1201 Morningside Drive, Manhattan Beach, CA 90266, USA 
(E-mail:  felixlev@hotmail.com)} \end{center}
\vskip 1em

{\it Abstract:}
\vskip 0.5em

In standard Poincare and anti de Sitter SO(2,3) invariant theories,
antiparticles are related to negative energy solutions of
covariant equations while independent positive energy unitary 
irreducible representations (UIRs) of the symmetry group
are used for describing both a particle and its antiparticle.
Such an approach 
cannot be applied in de Sitter SO(1,4) invariant theory. We argue 
that it would be more natural to require that (*) one UIR should 
describe a particle and its antiparticle simultaneously. This 
would automatically explain the existence of antiparticles and 
show that a 
particle and its antiparticle are different states of the same
object. If (*) is adopted then among the above groups only the
SO(1,4) one can be a candidate for constructing elementary particle
theory. It is shown that UIRs of the SO(1,4) group can be 
interpreted in the framework of (*) and cannot be interpreted 
in the standard way. By quantizing such UIRs and requiring 
that the energy should be positive in the Poincare approximation, 
we conclude that i) elementary particles can be only fermions.
It is also shown that ii) C invariance is not exact even in the 
free massive theory and iii) elementary particles cannot be neutral. 
This gives a natural explanation of the fact that all observed 
neutral states are bosons. 

\begin{flushleft} PACS: 11.30Cp, 11.30.Ly\end{flushleft}

\section{The statement of the problem}
\label{S1} 
 
In standard quantum theory the existence of antiparticles
is explained as follows. Each elementary particle is
described in two ways: i) by using a unitary irreducible 
representation (UIR) of the Poincare (or anti de Sitter) group; 
ii) by using a Poincare (or anti de Sitter) covariant 
equation. For each values of the mass and spin, there exist
two UIRs - with positive and negative energies, respectively.
At the same time, the corresponding covariant equation has
solutions with both, positive and negative energies.    
As noted by Dirac (see e.g. his Nobel
lecture \cite{DirNobel}), the existence of negative
energy solutions represents a difficulty which should be
resolved. In the standard approach, the solution is given
in the framework of quantization such that the
creation and annihilation operators for the antiparticle
have the usual meaning but they enter the quantum Lagrangian 
with the coefficients representing the negative energy
solutions. 

Such an approach has lead to impressive success in 
describing various experimental data. However, as noted by
Weinberg \cite{Wein1}, 'this is our aim in physics, not
just to describe nature, but to explain nature'. From this 
point of view, it seems unnatural that the covariant
equation describes the particle and antiparticle
simultaneously while UIRs for them are fully 
independent of each other. Moreover, UIRs with
negative energies are not used at all. 

The necessity to have negative energy solutions is
related to the implementation of the idea that the
creation or annihilation of an antiparticle can be treated,
respectively as the annihilation or creation of the 
corresponding particle with the negative energy. However,
since negative energies have no direct physical meaning in
the standard theory, this idea is implemented implicitly
rather than explicitly. 

The above program cannot be implemented if
the de Sitter (dS) group SO(1,4) is chosen as the 
symmetry group. For example, it is well known that
the dS Hamiltonian in UIRs has the spectrum in the 
interval $(-\infty,+\infty)$ (see e.g. Refs. 
\cite{Dobrev,Men,Mielke,Moy,lev1}).
Note also that in contrast with the anti de Sitter (AdS)
group SO(2,3), the dS one does not have a supersymmetric
generalization. In view of modern approaches to 
local quantum field theory (LQFT) in curved spacetime, 
the dS group cannot be the symmetry group 
since, from the standpoint of any local observer, the 
vacuum has a finite temperature and admits particle 
destruction and creation (see e.g. Refs. \cite{Narlikar}). 
For this and other reasons it was believed 
that the SO(1,4) group was not suitable for constructing 
elementary particle theory.

Although our approach considerably differs from that in 
Ref. \cite{Narlikar} and references therein
(see Ref. \cite{gravity} for a detailed discussion)
we come to the same conclusion (see Sect. \ref{S3})
that in the standard approach the dS group cannot be
a symmetry group. However, the standard approach can 
be modified in such a way (see below) that theories 
with the dS symmetry become consistent. The main
goal of the present paper is to investigate only the 
Poincare approximation where the standard physical 
intuition works. Some results of our investigations 
in the general case are mentioned in Sect. \ref{S8}.

It is well known that the group SO(1,4) is the symmetry 
group of the four-dimensional manifold in the five-dimensional 
space, defined by the equation 
\begin{equation}
 x_0^2-x_1^2-x_2^2-x_3^2-x_4^2=-R^2
\label{1}
\end{equation}
where a constant $R$ has the dimension of length.
The quantity $R^2$ is often written as $R^2=3/\Lambda$ where
$\Lambda$ is the cosmological constant. 
The nomenclature is such that $\Lambda < 0$ for  
the AdS symmetry while $\Lambda >0$ - for the dS one.
The recent astronomical data show that, although
$\Lambda$ is very small, it is probably positive 
(see e.g. Ref. \cite{Perlmutter}). For this reason the 
interest to dS theories has increased. Nevertheless, the 
existing difficulties have not been overcome 
(see e.g. Ref. \cite{Witten,Susskind}).

The fact that in the standard theory a particle and its
antiparticle are treated independently poses a problem
why they have equal masses, spins and lifetimes. The usual
explanation (see e.g. the standard textbooks \cite{AB})
is that this is a consequence of the CPT invariance.
Therefore if it appears that the masses of a particle and
its antiparticle were not equal, this would indicate the
violation of the CPT invariance. In turn, as shown in 
well-known works \cite{CPT}, any local Poincare invariant
quantum theory is automatically CPT invariant. 

Such an explanation seems to be not quite convincing.
Although at present there are no theories which explain
the existing data better than the standard model based on
LQFT, there is no guarantee that
the ultimate quantum theory will be necessarily local.
The modern theories aiming to unify all the known 
interactions (loop quantum gravity, noncommutative 
quantum theory, string theory etc.) do not
adopt the exact locality. Note also that the meaning
of time operator is not quite clear \cite{time} and, as 
it has been known already in 30th of the last century, when
quantum theory is combined with relativity, there is
no operator which has all the required properties of
the position operator (see e.g. Ref. \cite{NewtonWigner}).
In particular, it is not possible to localize an object 
with the
accuracy better than its Compton wave length. For this
and other reasons the quantity $x$ in the Lagrangian 
density $L(x)$ is only a parameter which becomes the 
space-time coordinate only in classical limit. 

Consider a model example when isotopic invariance
is exact (i.e. electromagnetic and weak interactions are 
absent). Then
the proton and the neutron have equal masses
and spins as a consequence of the fact that they belong 
to the same UIR of the isotopic group. In this example
the proton and the neutron are simply different states
of the same object - the nucleon, and the problem of why
they have equal masses and spins has a natural explanation. 

As shown in Ref. \cite{lev2}, in 
quantum theory based on a Galois field, Galois field analogs
of IRs of the AdS algebra so(2,3) have a property 
that a particle and its antiparticle are described by the 
same IR of the 
symmetry algebra. This automatically explains the existence 
of antiparticles and shows that a particle and its 
antiparticle represent different states of the same object. 
As argued in Ref. \cite{lev3,gravity}, 
the description of quantum theory in terms of Galois fields
is more natural than the standard description based on the
field of complex numbers. However, in the present paper we
consider only the standard approach but with the following 
modification. Instead of saying that UIRs (by definition)
describe elementary particles, we assume that

{\it Single-Representation Supposition:} In standard quantum 
theory, any unitary irreducible representations of the 
symmetry group or algebra should describe a particle and its 
antiparticle simultaneously.

With such a requirement, among the Poincare, AdS and 
dS groups,
only the latter can be a candidate for constructing the
elementary particle theory. Therefore, we have to investigate 
whether UIRs of the dS group are compatible 
with Single-Representation Supposition. In Sect. \ref{S2} we 
derive explicit
expressions for representation generators in UIRs and their
properties are discussed in Sect. \ref{S4}. 
In Sect. \ref{S3} the Poincare approximation is discussed and 
it is shown that UIRs of the dS group cannot be interpreted in the
standard way. In Sect.
\ref{S5} it is shown that the quantized version of UIR
is indeed compatible with Single-Representation Supposition 
and in the Poincare approximation the energy can be positive 
definite only for fermions. As shown in Sect. \ref{S6},
the antiparticle generators satisfy the correct commutation 
relations but the particle and antiparticle generators 
are different. As a consequence (see Sect. \ref{S7}),
elementary particles cannot be neutral, and even in the 
free massive dS invariant theory C invariance is only approximate.
Finally, Sect. \ref{S8} is discussion.        

\section{UIRs of the SO(1,4) group}
\label{S2}

As already noted, the de Sitter group SO(1,4) is the symmetry 
group of the four-dimensional manifold defined by Eq. (\ref{1}).
Elements of a map of the point $(0,0,0,0,R)$ (or $(0,0,0,0,-R)$) 
can be parametrized by the
coordinates $(x_0,x_1,x_2,x_3)$. If $R$ is very large then such a
map  proceeds
to Minkowski space and the action of the dS group on this map
--- to  the action of the Poincare group.

\begin{sloppypar}
The representation generators of the SO(1,4) group
$M^{ab}$ ($a,b=0,1,2,3,4$, $M^{ab}=-M^{ba}$) should satisfy the
commutation relations
\begin{equation}
[M^{ab},M^{cd}]=-i (\eta^{ac}M^{bd}+\eta^{bd}M^{as}-
\eta^{ad}M^{bc}-\eta^{bc}M^{ad})
\label{2}
\end{equation}
where $\eta^{ab}$ is the diagonal metric tensor such that
$\eta^{00}=-\eta^{11}=-\eta^{22}=-\eta^{33}=-\eta^{44}=1$.
\end{sloppypar}

In conventional quantum theory elementary particles 
are described by  UIRs of the symmetry group or IRs of its 
Lie algebra by
selfadjoint operators in Hilbert spaces. Usually the latter 
also are called UIRs
having in mind that the representation of the Lie algebra 
can be extended to the representation of the 
corresponding Lie group. We also will not discuss the
difference between Hermitian and selfadjoint operators.  

If one assumes that the role of the symmetry group is played by
the Poincare group then the representations are described by
ten generators --- six generators of the Lorentz group and four
components of the momentum operator. In the units $c=\hbar =1$ 
the former are
dimensionless while the latter has the dimension $(length)^{-1}$.
If however, the symmetry group is SO(1,4) (or SO(2,3)), then
all the generators in the units $c=\hbar =1$ are dimensionless. 

The main goal of this section is to derive explicit
expressions for the operators $M^{ab}$ in the case of
principal series of UIRs describing massive elementary particles.
Although there exists a wide literature on UIRs of the
SO(1,4) group (see e.g. Refs. 
\cite{Dix1,Tak,Hann,Str,Dobrev,Men,Mielke,Moy,lev1}), 
we did not succeed in finding these
expressions in the literature. For this reason we will
describe in detail the steps needed for derivation of
Eqs. (\ref{17}) and (\ref{18}).

The first complete mathematical classification of the
UIRs has been given in Ref. \cite{Dix1}, three well-known
realizations of the UIRs have been first considered in 
Ref. \cite{Tak}
and their physical context has been first discussed in 
Ref. \cite{Hann}.
It is well known that for classification of UIRs we should,
strictly speaking, consider not the group SO(1,4) 
itself but its
universal covering group. The investigation carried out in
Refs. \cite{Dix1,Tak,Hann,Str,Moy} has shown that this 
involves only replacement of the SO(3) group by its universal 
covering group SU(2). Since this procedure is well known,
we will work with the SO(1,4) group itself and 
follow a very elegant presentation for physicists in terms
of induced representations, given in the book \cite{Men}
(see also Refs. \cite{Dobrev,Str}). 

The elements of the SO(1,4) group will be described in the
block form
\begin{equation}
g=\left\|\begin{array}{ccc}
g_0^0&{\bf a}^T&g_4^0\\
{\bf b}&r&{\bf c}\\
g_0^4&{\bf d}^T&g_4^4
\end{array}\right\|\ 
\label{3}
\end{equation}
where 
\begin{equation}
\label{4}
{\bf a}=\left\|\begin{array}{c}a^1\\a^2\\a^3\end{array}\right\| \quad
{\bf b}^T=\left\|\begin{array}{ccc}b_1&b_2&b_3\end{array}\right\|
\quad r\in SO(3)
\end{equation}
(the subscript $^T$ means a transposed vector).

UIRs of the SO(1,4) 
group are induced from UIRs of the subgroup $H$ defined
as follows \cite{Men}. Each element of $H$ can be uniquely
represented as a product of elements of the subgroups
SO(3), $A$ and ${\bf T}$: $h=r\tau_A{\bf a}_{\bf T}$ where 
\begin{equation}
\tau_A=\left\|\begin{array}{ccc}
cosh(\tau)&0&sinh(\tau)\\
0&1&0\\
sinh(\tau)&0&cosh(\tau)
\end{array}\right\|\ \quad
{\bf a}_{\bf T}=\left\|\begin{array}{ccc}
1+{\bf a}^2/2&-{\bf a}^T&{\bf a}^2/2\\
-{\bf a}&1&-{\bf a}\\
-{\bf a}^2/2&{\bf a}^T&1-{\bf a}^2/2
\end{array}\right\|\ 
\label{5}
\end{equation}

The subgroup $A$ is one-dimensional and the three-dimensional
group ${\bf T}$ is the dS analog of the conventional
translation group (see e.g. Ref. \cite{Men,Mielke}). We hope it 
should not cause misunderstandings when 1 is used in its
usual meaning and when to denote the unit element of the
SO(3) group. It should also be clear when $r$ is a true
element of the SO(3) group or belongs to the SO(3) subgroup
of the SO(1,4) group. 

Let $r\rightarrow \Delta(r;{\bf s})$ be a UIR of the group
SO(3) with the spin ${\bf s}$ and 
$\tau_A\rightarrow exp(i\mu\tau)$ be a
one-dimensional UIR of the group $A$, where $\mu$ is a real
parameter. Then UIRs of the group $H$ used for inducing to
the SO(1,4) group, have the form
\begin{equation}
\Delta(r\tau_A{\bf a}_{\bf T};\mu,{\bf s})=
exp(i\mu\tau)\Delta(r;{\bf s})
\label{6}
\end{equation} 
We will see below that $\mu$ has the meaning of the dS
mass and therefore UIRs of the SO(1,4) group are
defined by the mass and spin, by analogy with UIRs
in Poincare invariant theory.

Let $G$=SO(1,4) and $X=G/H$ be the factor space (or
coset space) of $G$ over $H$. The notion of the factor 
space is well known (see e.g. Ref. \cite{Naimark,Dobrev,Men}).
Each element $x\in X$ is a class containing the
elements $x_Gh$ where $h\in H$, and $x_G\in G$ is a
representative of the class $x$. The choice of
representatives is not unique since if $x_G$ is
a representative of the class $x\in G/H$ then
$x_Gh_0$, where $h_0$ is an arbitrary element
from $H$, also is a representative of the same 
class. It is well known that $X$ can be treated 
as a left $G$ space. This means that if $x\in X$
then the action of the group $G$ on $X$ can be
defined as follows: if $g\in G$ then $gx$ is a
class containing $gx_G$ (it is easy to verify
that such an action is correctly defined). 
Suppose that the choice of representatives
is somehow fixed. Then $gx_G=(gx)_G(g,x)_H$
where $(g,x)_H$ is an element of $H$. This
element is called a factor.

The explicit form of the generators $M^{ab}$
depends on the choice of representatives in
the space $G/H$. As explained in several
papers devoted to UIRs of the SO(1,4) group
(see e.g. Ref. \cite{Men}), to obtain
the  possible closest analogy between UIRs of
the SO(1,4) and Poincare groups, one should proceed
as follows. Let ${\bf v}_L$ be a representative 
of the Lorentz group in the factor space SO(1,3)/SO(3)
(strictly speaking, we should consider $SL(2,C)/SU(2)$).
This space can be represented as the well known velocity
hyperboloid with the Lorentz invariant measure
\begin{equation}
d\rho({\bf v})=d^3{\bf v}/v_0
\label{7}
\end{equation}
where $v_0=(1+{\bf v}^2)^{1/2}$. Let $I\in SO(1,4)$ be a
matrix which formally has the same form as
the metric tensor $\eta$. One can show 
(see e.g. Ref. \cite{Men} for details) that 
$X=G/H$ can be represented as a union of three
spaces, $X_+$, $X_-$ and $X_0$ such that 
$X_+$ contains classes ${\bf v}_Lh$, $X_-$
contains classes ${\bf v}_LIh$ and $X_0$ is of
no interest for UIRs describing elementary particles 
since it has measure zero relative to the spaces
$X_+$ and $X_-$.

As a consequence, the space of UIR
of the SO(1,4) group can be implemented as follows.  
If $s$ is the spin of the particle under 
consideration, then we
use $||...||$ to denote the norm in the space of 
UIR of the group SU(2) with the spin $s$. 
Then the space of UIR is the space of 
functions $\{f_1({\bf v}),f_2({\bf v})\}$ on
two Lorentz hyperboloids with the range in the space of
UIR of the group SU(2) with the spin $s$ and such that
\begin{equation}
\int\nolimits [||f_1({\bf v})||^2+
||f_2({\bf v})||^2]d\rho({\bf v}) <\infty
\label{8}
\end{equation}

We see that, in contrast with UIRs of the Poincare 
group (and AdS one), where UIRs are implemented on
one Lorentz hyperboloid, UIRs of the dS group can be
implemented only on two Lorentz hyperboloids, $X_+$
and $X_-$. Even this
fact (which is well known) is a strong indication 
that UIRs of the dS group might
have a natural interpretation in the framework of
Single-Representation Supposition (to the best of our 
knowledge, this possibility has not been considered in 
the literature).

In the case of Poincare and AdS groups, the positive
energy UIRs are implemented on an analog of $X_+$ and 
negative energy UIRs - on an analog of $X_-$. Since the 
Poincare and AdS groups
do not contain elements transforming these spaces
to one another, the positive and negative energy UIRs 
are fully independent. At the same time, the dS 
group contains 
such elements (e.g. $I$ \cite{Men,Mielke}) and for 
this reason its UIRs cannot be implemented only on 
one hyperboloid. 

A general construction of the operators $M^{ab}$ is as 
follows. We first define right
invariant measures on $G=SO(1,4)$ and $H$. It is well
known (see e.g. Ref. \cite{Naimark}) that for 
semisimple Lie groups (which is the case for the
dS group), the right invariant measure is
simultaneously the left invariant one. At the same
time, the right invariant measure $d_R(h)$ on $H$ is not
the left invariant one, but has the property
$d_R(h_0h) = \Delta(h_0)d_R(h)$, where the number
function $h\rightarrow \Delta(h)$ on $H$ is called
the module of the group $H$. It is easy to show 
\cite{Men} that
\begin{equation}
\Delta(r\tau_A{\bf a}_{\bf T})=exp(-3\tau)
\label{9}
\end{equation}
Let $d\rho(x)$ be a measure on $X=G/H$ compatible with
the measures on $G$ and $H$ \cite{Men} and let the
representation space be implemented as the space of
functions $\varphi(x)$ on $X$ with the range in the
space of UIR of the SU(2) group such that
\begin{equation}
\int\nolimits ||\varphi(x)||^2d\rho(x) <\infty
\label{10}
\end{equation}
Then the action of the representation operator $U(g)$
corresponding to $g\in G$ is defined as \cite{Men}
\begin{eqnarray}
U(g)\varphi(x)=[\Delta((g^{-1},x)_H)]^{-1/2}
\Delta((g^{-1},x)_H;\mu,{\bf s})^{-1}\varphi(g^{-1}x)
\label{11}
\end{eqnarray}
One can verify that this expression indeed defines
a unitary representation. Its irreducibility can be
proved in several ways (see e.g. Ref. \cite{Men}).

As noted above, one can use the realization of the
space $X$ as the union of $X_+$ and $X_-$ and then
the representation space can be realized as in Eq.
(\ref{8}). Since we are interested in calculating the
explicit form of representation generators, it is
sufficient for this purpose to consider only elements
of $g\in G$ in an infinitely small vicinity of the unit
element of the dS group. In that case one can calculate
the action of representation operators on functions
having the carrier in $X_+$ and $X_-$ separately.
Namely, as follows from Eq. (\ref{11}), for such 
$g\in G$, one has to find the decompositions
\begin{equation}
g^{-1}{\bf v}_L={\bf v}'_Lr'(\tau')_A({\bf a}')_{\bf T}
\label{12}
\end{equation}
and
\begin{equation}
g^{-1}{\bf v}_LI={\bf v}"_LIr"(\tau")_A({\bf a}")_{\bf T}
\label{13}
\end{equation}
where $r',r"\in SO(3)$. 
In this expressions it is sufficient to consider
only the elements of $H$ belonging to an 
infinitely small vicinity of the unit element. 

The problem of choosing representatives in the factor
space SO(1,3)/SO(3) (or SL(2.C)/SU(2)) is well known in 
the standard theory. The most usual choice 
is such that ${\bf v}_L\in SO(1,4)$ is represented by
the matrix
\begin{equation}
{\bf v}_L=\left\|\begin{array}{ccc}
v_0&{\bf v}^T&0\\
{\bf v}&1+{\bf v}{\bf v}^T/(v_0+1)&0\\
0&0&1
\end{array}\right\|\ 
\label{14}
\end{equation}

As follows from Eqs. (\ref{6}) and (\ref{11}), there is
no need to know the expressions for $({\bf a}')_{\bf T}$
and $({\bf a}")_{\bf T}$ in Eqs. (\ref{12}) and (\ref{13}).
We can use the fact \cite{Men} that if $e$ is the 
five-dimensional vector with the components 
$(e^0=1,0,0,0,e^4=-1)$ and $h=r\tau_A{\bf a}_{\bf T}$,
then $he=exp(-\tau)e$ regardless of the elements 
$r\in SO(3)$ and ${\bf a}_{\bf T}$. This makes it possible
to easily calculate $({\bf v}'_L,{\bf v}"_L,(\tau')_A,(\tau")_A)$
in Eqs. (\ref{12}) and (\ref{13}). Then one can calculate
$(r',r")$ in these expressions by using the fact that the
SO(3) parts of the matrices $({\bf v}'_L)^{-1}{\bf v}_L$
and $({\bf v}"_L)^{-1}{\bf v}_L$ are equal to $r'$ and
$r"$, respectively.

The relation between the operators $U(g)$ and $M^{ab}$ is
as follows. Let $L_{ab}$ be the basis elements of the Lie
algebra of the dS group. These are the matrices with the
elements
\begin{equation}
(L_{ab})_d^c=\delta_d^c\eta_{bd}-\delta_b^c\eta_{ad}
\label{15}
\end{equation}
They satisfy the commutation relations
\begin{equation}
[L_{ab},L_{cd}]=\eta_{ac}L_{bd}-\eta_{bc}L_{ad}-
\eta_{ad}L_{bc}+\eta_{bd}L_{ac}
\label{16}
\end{equation}
Comparing Eqs. (\ref{2}) and (\ref{16}) it is easy
to conclude that the $M^{ab}$ should be the
representation operators of $-iL^{ab}$. Therefore
if $g=1+\omega_{ab}L^{ab}$, where a sum over repeated indices 
is assumed and the $\omega_{ab}$ are
such infinitely small parameters that 
$\omega_{ab}=-\omega_{ba}$ then $U(g)=1+i\omega_{ab}M^{ab}$.

\begin{sloppypar}
We are now in position to write down the final expressions
for the representation generators.
The  explicit  calculation  shows  that  the  action  of the
generators on functions with the carrier in
$X_+$ has the form
\begin{eqnarray}
&&{\bf M}^{(+)}=l({\bf v})+{\bf s},\quad {\bf N}^{(+)}=-i v_0
\frac{\partial}{\partial {\bf v}}+\frac{{\bf s}\times {\bf v}}
{v_0+1}, \nonumber\\
&& {\bf B}^{(+)}=\mu {\bf v}+i [\frac{\partial}{\partial {\bf v}}+
{\bf v}({\bf v}\frac{\partial}{\partial {\bf v}})+\frac{3}{2}{\bf v}]+
\frac{{\bf s}\times {\bf v}}{v_0+1},\nonumber\\
&& M_{04}^{(+)}=\mu v_0+i v_0({\bf v}
\frac{\partial}{\partial {\bf v}}+\frac{3}{2})
\label{17}
\end{eqnarray}
where ${\bf M}=\{M^{23},M^{31},M^{12}\}$,
${\bf N}=\{M^{01},M^{02},M^{03}\}$,
${\bf B}=-\{M^{14},M^{24},M^{34}\}$, ${\bf s}$ is the spin operator,
and ${\bf l}({\bf v})=-i{\bf v}
\times \partial/\partial {\bf v}$.
At the same time, the action of the generators on 
functions with the carrier 
in $X_-$ is given by
\begin{eqnarray}
&&{\bf M}^{(-)}=l({\bf v})+{\bf s},\quad {\bf N}^{(-)}=-i v_0
\frac{\partial}{\partial {\bf v}}+\frac{{\bf s}\times {\bf v}}
{v_0+1}, \nonumber\\
&& {\bf B}^{(-)}=-\mu {\bf v}-i [\frac{\partial}{\partial {\bf v}}+
{\bf v}({\bf v}\frac{\partial}{\partial {\bf v}})+\frac{3}{2}{\bf v}]-
\frac{{\bf s}\times {\bf v}}{v_0+1},\nonumber\\
&& M_{04}^{(-)}=-\mu v_0-i v_0({\bf v}
\frac{\partial}{\partial {\bf v}}+\frac{3}{2})
\label{18}
\end{eqnarray}
\end{sloppypar}

\section{Poincare approximation}
\label{S3}

Consider first the case when UIRs of the dS group can
be contracted to UIRs of the Poincare group. A
general notion of contraction has been developed in 
Ref. \cite{IW}. In our case it can be performed
as follows. We assume that $\mu > 0$, denote
$m=\mu /R$ and formally consider the limit 
$R\rightarrow \infty$ when $\mu\rightarrow \infty$
but $\mu /R$ is finite. We will consider the limits
for Eqs. (\ref{17}) and (\ref{18}) separately.

In the case of Eq. (\ref{17}) we denote
${\bf P}={\bf B}/R$ and $E=M_{04}/R$.
Then in the above limit we obtain a standard 
representation of the
Poincare group for a particle with the mass $m$ such 
that ${\bf P}=m{\bf v}$ is the particle momentum
and $E=mv_0$ is the particle energy. In that case
the generators of the Lorentz group have the same form
for the Poincare and dS groups. However, if the same 
procedure is applied to Eq. (\ref{18})
then the quantity ${\tilde E}=M_{04}/R$ becomes
negative and therefore the problem arises whether
${\tilde E}$ can be identified with the standard energy
$E$.

The following important observation is in order. If we
assume that the dS symmetry is more fundamental than
the Poincare one then the limit $R\rightarrow\infty$ 
{\it should not} be actually taken since in this case the
dS symmetry will be lost and the preceding consideration
will become useless. However, in the framework of the
dS invariance we can consider Poincare invariance as 
the approximate symmetry when $R$ is very large but 
$R\neq\infty$. The situation is analogous to that when 
nonrelativistic theory is formally 
treated as a special case of relativistic one in 
the limit $c\rightarrow\infty$ and classical theory 
is treated as a special case of quantum one in
the limit $\hbar\rightarrow 0$ but the limits are
not actually taken. Summarizing these remarks, we prefer
the term 'Poincare approximation' rather than
'Poincare limit'. {\it The term 'Poincare approximation' 
will always imply that $R$ is very large but finite.}    

In the standard interpretation of UIRs
the following requirements should be satisfied:
\begin{itemize}
\item {\it Standard-Interpretation Requirements.} 
Each element of the full representation space 
represents a possible physical state for the given 
elementary particle. The representation describing
a system of $N$ free elementary particles is the tensor
product of the corresponding single-particle
representations.
\end{itemize}

Recall that the generators of the tensor product are
given by sums of the corresponding single-particle
generators. For example, if $M_{04}^{(1)}$ is the
operator $M_{04}$ for particle 1 and $M_{04}^{(2)}$ is 
the operator $M_{04}$ for particle 2 then the operator
$M_{04}$ for the free system $\{12\}$ is given by
$M_{04}^{(12)}=M_{04}^{(1)}+M_{04}^{(2)}$. Here it is assumed
that the action of the operator $M_{04}^{(j)}$ ($j=1,2$)
in the two-particle space is defined as follows. It
acts according to Eq. (\ref{17}) or (\ref{18}) over its
respective variables while over the variables of the
other particle it acts as the identity operator. 
 
It is well known (see e.g. Ref. 
\cite{Men,Dobrev,Moy,Mielke})
that the dS group contains elements (e.g. $I$)
such that the corresponding representation operator
transforms eigenstates of $M_{04}$ with
the positive eigenvalues to the eigenstates of the
same operator with the negative eigenvalues and 
{\it vice versa}. Therefore the problem arises
whether the operator $M_{04}$ in the single-particle
UIRs can be treated as the dS analog of the energy and 
what is the relation between $M_{04}$ and the standard 
single-particle energy $E$. Let us stress that {\it if
Poincare invariance is treated as a special case of
the dS one then the quantity $E$ cannot be defined 
independently and should somehow be expressed in terms 
of the operators $M^{ab}$}.
One could try to remedy the 
standard interpretation as follows.
\begin{itemize}
\item Assume that in the Poincare
approximation the standard energy should be defined as 
\begin{equation}
E = \pm M_{04}/R
\label{Ref2}
\end{equation} 
where the plus sign should be taken for the states with 
the carrier in $X_+$ and as the minus sign ---
for the states with the carrier in $X_-$. Then the 
energy will always be positive definite. 
\item One might say that the choice of 
the energy sign is only a matter of convention. 
Indeed, to measure the energy of a particle with
the mass $m$ one has to measure its momentum ${\bf p}$
and then the energy can be defined not only as 
$(m^2+{\bf p}^2)^{1/2}$ but also as $-(m^2+{\bf p}^2)^{1/2}$.
In that case the standard energy in the Poincare approximation
could be defined as 
\begin{equation}
E = M_{04}/R
\label{Ref2B}
\end{equation} 
regardless of whether the carrier of the given state is
in $X_+$ or $X_-$.
\end{itemize}

It is easy to see that either of the above
possibilities is incompatible with Standard-Interpretation
Requirements. Consider, for example, a system of two free 
particles in the Poincare approximation. 
Then with a high accuracy the operators $M_{04}/R$ and 
${\bf B}/R$ can be chosen diagonal simultaneously. 

Let us first assume that the energy should be treated
according to Eq. (\ref{Ref2B}). Then a system of two free 
particles with the equal
masses can have the same quantum numbers as the
vacuum (for example, if the first particle has the
energy $E_0=(m^2+{\bf p}^2)^{1/2}$ and momentum 
${\bf p}$ while the second
one has the energy $-E_0$ and the momentum $-{\bf p}$)
what obviously contradicts experiment.
For this and other reasons it is well known that in the
Poincare invariant theory the particles
should have the same energy sign. Analogously,
if the single-particle energy is treated according to 
Eq. (\ref{Ref2}) and one requires that the two-body 
energy is the sum of single-particle energies then 
Eq. (\ref{Ref2}) for the two-body system will not be 
satisfied. 

We conclude that UIRs of the dS algebra cannot be 
interpreted in the standard way since such an
interpretation is physically meaningless even in
the Poincare approximation. Although our approach 
considerably differs from LQFT in curved
spacetime (see Sect. \ref{S1}), this conclusion is 
in agreement with that in Ref. \cite{Narlikar} and 
references therein. 

In the framework of Single-representation supposition, 
one could try to
interpret the operators (\ref{17}) as those describing
a particle while the operators (\ref{18}) as those
describing the corresponding antiparticle. This will
be done in the subsequent sections and we will see that
for quantized operators the energy can be interpreted 
as $E = M_{04}/R$.

It is clear that the above contraction procedure is
valid only if $\mu\neq 0$. Therefore if we accept that
Poincare invariant theory is a special case of the
dS invariant one, the problem arises how to describe
particles which in Poincare invariant theory are 
strictly massless (and whether such particles can
exist). Mensky has suggested \cite{Men} that massless 
particles should be described by UIRs corresponding
to the additional series with $-i\mu=1/2$. This 
problem requires further study.

\section{Properties of representation generators}
\label{S4}

We now return to the general case 
when the quantity $R$ is not necessarily large.
Let us first compare Eqs. (\ref{17}) and (\ref{18}). As
follows from Eq. (\ref{2}), if a set 
$M_{ab}$ satisfies the correct commutation relations, 
the same is true for the set obtained from $M_{ab}$ by
changing the sign of those operators where $a=4$ or 
$b=4$ (the operator $M_{44}$ is identical zero since
$M_{ab}=-M_{ba}$). Therefore if one wants to verify
that the operators (\ref{17}) and (\ref{18}) 
satisfy the conditions (\ref{2}), it is sufficient to
verify this either for (\ref{17}) or (\ref{18}).

It is obvious that the operators obtained from (\ref{17})
or (\ref{18}) by the transformation 
$\mu\rightarrow -\mu$ satisfy the conditions (\ref{2})
if the original operators satisfy these conditions.
Let us now apply the following transformation. First
change the sign of $\mu$ and then change the sign of
those operators $M_{ab}$ where $a=4$ or $b=4$. Then we
obtain a set of operators satisfying
Eq. (\ref{2}) if the original set satisfies Eq. (\ref{2}).
If such a transformation is applied to  (\ref{17}),
we obtain the following set of operators
\begin{eqnarray}
&&{\bf M}'=l({\bf v})+{\bf s},\quad {\bf N}'=-i v_0
\frac{\partial}{\partial {\bf v}}+\frac{{\bf s}\times {\bf v}}
{v_0+1}, \nonumber\\
&& {\bf B}'=\mu {\bf v}-i [\frac{\partial}{\partial {\bf v}}+
{\bf v}({\bf v}\frac{\partial}{\partial {\bf v}})+\frac{3}{2}{\bf v}]-
\frac{{\bf s}\times {\bf v}}{v_0+1},\nonumber\\
&& M_{04}'=\mu v_0-i v_0({\bf v}
\frac{\partial}{\partial {\bf v}}+\frac{3}{2})
\label{19}
\end{eqnarray}

By using Eqs. (\ref{7}) and (\ref{8}), one can directly 
verify that the operators (\ref{17}), (\ref{18}) and 
(\ref{19}) are 
Hermitian if the scalar product in the space of UIR
is defined as follows. Since the functions $f_1({\bf v})$
and $f_2({\bf v})$ in Eq. (\ref{8}) have the range in the
space of UIR of the group SU(2) with the spin $s$, we can
replace them by the sets of functions $f_1({\bf v},j)$ and
$f_2({\bf v},j)$, respectively, where $j=-s,-s+1...s$.
Moreover, we can combine these functions into one function
$f({\bf v},j,\epsilon)$ where the variable $\epsilon$ can
take only two values, say +1 or -1, for the components
having the carrier in $X_+$ or $X_-$, respectively.
If now $\varphi({\bf v},j,\epsilon)$ and
$\psi({\bf v},j,\epsilon)$ are two elements of our 
Hilbert space, their scalar product is defined as
\begin{equation}
(\varphi,\psi)=\sum_{j,\epsilon}\int\nolimits 
\varphi({\bf v},j,\epsilon)^*\psi({\bf v},j,\epsilon)
d\rho({\bf v}
\label{20})
\end{equation}
where the subscript $^*$ applied to scalar functions
means the usual complex conjugation.

At the same time, we use $^*$ to denote the
operator adjoint to a given one. Namely, if $A$ is the
operator in our Hilbert space then $A^*$ means the
operator such that
\begin{equation}
(\varphi,A\psi)=(A^*\varphi,\psi)
\label{21}
\end{equation}
for all such elements $\varphi$ and $\psi$ that the left
hand side of this expression is defined.  
 
Even in the case of the operators (\ref{17}), (\ref{18})
and (\ref{19}), we can formally treat them as integral 
operators with some kernels.
Namely, if $A\varphi=\psi$, we can treat this relation as 
\begin{equation}
\sum_{j',\epsilon'}\int\nolimits 
A({\bf v},j,\epsilon;{\bf v}',j',\epsilon')
\varphi({\bf v}',j',\epsilon')d\rho({\bf v}')=
\psi({\bf v},j,\epsilon) 
\label{22}
\end{equation} 
where in the general case the kernel
$A({\bf v},j,\epsilon;{\bf v}',j',\epsilon')$ of the
operator $A$ is a distribution.

As follows from Eqs. (\ref{7}), (\ref{21}) and (\ref{22}),
if $B=A^*$ then the relation between the kernels of these
operators is as follows:
\begin{equation}
B({\bf v},j,\epsilon;{\bf v}',j',\epsilon')=
A({\bf v}',j',\epsilon';{\bf v},j,\epsilon)^*
\label{23}
\end{equation}
In particular, if the operator $A$ is Hermitian then
\begin{equation}
A({\bf v},j,\epsilon;{\bf v}',j',\epsilon')^*=
A({\bf v}',j',\epsilon';{\bf v},j,\epsilon)
\label{24}
\end{equation}

As follows from Eq. (\ref{24}), if the operator
$A$ is Hermitian, and its kernel is real then the
kernel is symmetric, i.e.
\begin{equation}
A({\bf v},j,\epsilon;{\bf v}',j',\epsilon')=
A({\bf v}',j',\epsilon';{\bf v},j,\epsilon)
\label{25}
\end{equation}
In particular, this property is satisfied for the
operators $\mu v_0$ and $\mu {\bf v}$ in Eqs.
(\ref{17}, (\ref{18}) and (\ref{19}). At the same 
time, the operators
\begin{equation}
l({\bf v}),\quad -i v_0\frac{\partial}{\partial {\bf v}},
\quad -i [\frac{\partial}{\partial {\bf v}}+
{\bf v}({\bf v}\frac{\partial}{\partial 
{\bf v}})+\frac{3}{2}{\bf v}],
\quad -i v_0({\bf v}
\frac{\partial}{\partial {\bf v}}+\frac{3}{2})
\label{26}
\end{equation}
which are present in Eqs. (\ref{17}, (\ref{18}) and 
(\ref{19}),
are Hermitian but have imaginary kernels. Therefore,
as follows from Eq. (\ref{24}), their kernels are 
antisymmetric:
\begin{equation}
A({\bf v},j,\epsilon;{\bf v}',j',\epsilon')=-
A({\bf v}',j',\epsilon';{\bf v},j,\epsilon)
\label{27}
\end{equation}
Note also that the operators considered in this
paragraph do not depend on the spin and are present 
in Eqs. (\ref{17}, (\ref{18}) and (\ref{19}) for 
particles with
arbitrary spins. At the same time, the spin operator
is obviously different for particles with different
spins. This question will be considered in Sect. \ref{S6}.

\section{Quantization of UIRs}
\label{S5}

In standard approach to quantum theory, the operators of 
physical quantities act in the Fock space of the 
given system. Suppose that the system consists
of free particles and their antiparticles.
Strictly speaking, in our approach it is not clear yet
what should be treated as a particle or antiparticle.
The considered UIRs of the dS group describe objects
such that $({\bf v}, j, \epsilon)$ is the full set of
their quantum numbers. Therefore we can define the
annihilation and creation operators 
$(a({\bf v},j,\epsilon),a({\bf v},j,\epsilon)^*)$
for these objects. If the 
operators satisfy the
anticommutation relations then we require that
\begin{equation}
\{a({\bf v},j,\epsilon),a({\bf v}',j',\epsilon')^*\}=
\delta_{jj'}\delta_{\epsilon\epsilon'}v_0
\delta^{(3)}({\bf v}-{\bf v}')
\label{28}
\end{equation}
while in the case of commutation relations
\begin{equation}
[a({\bf v},j,\epsilon),a({\bf v}',j',\epsilon')^*]=
\delta_{jj'}\delta_{\epsilon\epsilon'}v_0
\delta^{(3)}({\bf v}-{\bf v}')
\label{29}
\end{equation}
In the first case, any two $a$-operators or any two
$a^*$ operators anticommute with each other while
in the second case they commute with each other. 

The problem of second quantization of representation 
operators can
now be formulated as follows. Let $(A_1,A_2....A_n)$ 
be representation
generators describing UIR of the dS group. One should
replace them by operators acting in the Fock space 
such that the commutation relations between their
images in the Fock space are the same as for original
operators (in other words, we should have a homomorphism
of Lie algebras of operators acting in the space of UIR
and in the Fock space). We can also require that our 
map should be compatible with the Hermitian
conjugation in both spaces. It is easy to verify that
a possible solution satisfying all the requirements is
as follows. If the operator $A$ in the space of UIR
has the kernel $A({\bf v},j,\epsilon;{\bf v}',j',\epsilon')$
then the image of A in the Fock space is the operator 
\begin{equation}
A_F=\sum_{j,\epsilon,j',\epsilon'}\int\nolimits\int\nolimits
A({\bf v},j,\epsilon;{\bf v}',j',\epsilon')
a({\bf v},j,\epsilon)^*a({\bf v}',j',\epsilon')
d\rho({\bf v})d\rho({\bf v}')
\label{30}
\end{equation}
The commutation relations in the Fock space will be 
preserved regardless of whether the $(a,a^*)$ operators 
satisfy commutation
or anticommutation relations.

We now require that in the Poincare approximation the energy
should be positive definite. Recall that the operators 
(\ref{17},\ref{18}) act on their respective subspaces or in
other words, they are diagonal in the quantum 
number $\epsilon$.

Suppose that $\mu> 0$ and consider the quantized operator
corresponding to the dS energy $M_{04}$ in Eq. (\ref{17}).
In the Poincare approximation, 
$M_{04}^{(+)}=\mu v_0$ is fully
analogous to the standard free energy and therefore, as
follows from Eq. (\ref{30}), its quantized form is
\begin{equation}
(M_{04}^{(+)})_F=\mu\sum_{j}\int\nolimits v_0
a({\bf v},j,1)^*a({\bf v},j,1)d\rho({\bf v})
\label{31}
\end{equation} 
This expression is fully analogous to the standard
quantized Hamiltonian if we assume that the vacuum state
$\Phi_0$ satisfies the requirement 
\begin{equation}
a({\bf v},j,1)\Phi_0=0\quad \forall\,\, {\bf v},j
\label{vac1}
\end{equation}
In this case $a({\bf v},j,1)$ has the meaning of
the annihilation operator, $a({\bf v},j,1)^*$ has the
meaning of the creation operator, and 
$a({\bf v},j,1)^*\Phi_0$ has the meaning of the
one-particle state.

Consider now the operator
$M_{04}^{(-)}$. In the Poincare approximation 
its quantized form is
\begin{equation}
(M_{04}^{(-)})_F==-\mu\sum_{j}\int\nolimits v_0
a({\bf v},j,-1)^*a({\bf v},j,-1)d\rho({\bf v})
\label{32}
\end{equation} 
Therefore, if, by analogy with Eq. (\ref{vac1}), one 
requires that
\begin{equation}
a({\bf v},j,-1)\Phi_0=0\quad \forall\,\, {\bf v},j
\label{vac2}
\end{equation}
then the operator $(M_{04}^{(-)})_F$ will be negative
definite, what is unacceptable.

Therefore the operators $a({\bf v},j,-1)$ and 
$a({\bf v},j,-1)^*$ are "nonphysical": 
$a({\bf v},j,-1)$ is the operator of object's
annihilation with the negative energy, and 
$a({\bf v},j,-1)^*$ is the operator of object's
creation with the negative energy.

We will interpret the operator $(M_{04}^{(-)})_F$ as 
that related to antiparticles. As already noted, in 
the standard approach, the annihilation and creation
operators for antiparticles enter the quantum Lagrangian
with the coefficients describing negative energy solutions
of the corresponding covariant equation. This is an
implicit implementation of the idea that  
the creation or annihilation of an antiparticle can be 
treated, respectively as the annihilation or creation of 
the corresponding particle with the negative energy.
In our case this idea can be implemented explicitly. 

Instead of the operators $a({\bf v},j,-1)$ and 
$a({\bf v},j,-1)^*$, we define new operators
$b({\bf v},j)$ and $b({\bf v},j)^*$. If 
$b({\bf v},j)$ is treated as the "physical" operator
of antiparticle annihilation then, according to the
above idea, it should be proportional to  
$a({\bf v},-j,-1)^*$. Analogously, if $b({\bf v},j)^*$
is the "physical" operator of antiparticle creation,
it should be proportional to $a({\bf v},-j,-1)$.
Therefore
\begin{equation}
b({\bf v},j)=a({\bf v},-j,-1)^*/\eta (j)\quad 
b({\bf v},j)^*= a({\bf v},-j,-1)/\eta (j)^*
\label{33}
\end{equation}
where $\eta (j)$ is a phase factor such that
\begin{equation}
|\eta (j)|=1
\label{34}
\end{equation}

Since we treat $b({\bf v},j)$ as the annihilation
operator and $b({\bf v},j)^*$ as the creation one,
instead of Eq. (\ref{vac2}) we should require that
the vacuum condition should read
\begin{equation}
b({\bf v},j)\Phi_0=0\quad \forall\,\, {\bf v},j,
\label{vac3}
\end{equation}
in the case of anticommutation relations  
\begin{equation}
\{b({\bf v},j),b({\bf v}',j')^*\}=
\delta_{jj'}v_0 \delta^{(3)}({\bf v}-{\bf v}'),
\label{36}
\end{equation}
and in the case of commutation relations
\begin{equation}
[b({\bf v},j),b({\bf v}',j')^*]=
\delta_{jj'}v_0 \delta^{(3)}({\bf v}-{\bf v}')
\label{comm}
\end{equation}

Consider first the case when the operators
$a({\bf v},j,\epsilon)$ satisfy the anticommutation
relations (\ref{28}). By using Eq. (\ref{33}) one
can express the operators $a({\bf v},j,-1)$ in terms
of the operators $b({\bf v},j)$. Then it follows from
the condition (\ref{34}) that the the operators 
$b({\bf v},j)$ indeed satisfy Eq. (\ref{36}). 

Consider now the case when the operators
$a({\bf v},j,\epsilon)$ satisfy the commutation
relations (\ref{29}). We can again use Eq. (\ref{33}) 
to express the operators $a({\bf v},j,-1)$ in terms
of the operators $b({\bf v},j)$. However, it now
follows from the condition (\ref{34}) that the
operators $b({\bf v},j)$ do not satisfy Eq. 
(\ref{comm})) (they satisfy the equation obtained
from Eq. (\ref{comm}) by changing the sign of the
r.h.s.). 

These results show that only in the
case of fermions our construction might be 
consistent. To see whether this is the case, we should 
express the operator (\ref{32}) in terms of the
operators $b({\bf v},j)$. By using Eqs. (\ref{33}) and 
(\ref{34}), we can rewrite Eq. (\ref{32}) as
\begin{equation}
(M_{04}^{(-)})_F==-\mu\sum_{j}\int\nolimits v_0
b({\bf v},j)b({\bf v},j)^*d\rho({\bf v})
\label{35}
\end{equation}
Now we have a situation fully analogous to that
described in various textbooks 
(see e.g. Ref. \cite{AB}) for quantizing the
electron-positron field. It is well known
that the only way to ensure
the positive definiteness is to require that the
operators $b({\bf v},j)$ and $b({\bf v},j)^*$
should satisfy the anticommutation relations
(\ref{36}). Then we can rewrite Eq. (\ref{35}) as
\begin{equation}
(M_{04}^{(-)})_F==\mu\sum_{j}\int\nolimits v_0
b({\bf v},j)^*b({\bf v},j)d\rho({\bf v})+C
\label{37}
\end{equation}
where $C$ is some indefinite constant. It can be
eliminated by requiring that all quantized
operators should be written in the normal form or
by using another prescriptions. The existence of
infinities in the standard approach is the well known
problem and we will not discuss it. 

Our conclusion is as follows: 
\begin{itemize}
\item {\it Statement 1:} The requirement
that the Hamiltonian should be positive definite in
the Poincare approximation, can be satisfied only for 
fermions.
\end{itemize}   

\section{Antiparticle sector}
\label{S6}

In the preceding section we argued that the 
$(b,b^*)$ operators are physical operators
describing annihilation and creation of antiparticles.
However, the proof of this statement has been
given only in the Poincare approximation. 
To prove the statement in the general case we must show 
that the quantized operators (\ref{18}) written in 
terms of the $(b,b^*)$ operators satisfy the correct 
commutation relations (\ref{2}). We can use 
Eqs. (\ref{33}), (\ref{34}) to express the operators
$b({\bf v},j)$ in terms of $a({\bf v},j, -1)$, and,  
since we are now interested only in the case of 
anticommutation relations, we assume that Eq.
(\ref{36}) is satisfied. 

Consider first the operators $-\mu v_0$ and $-\mu {\bf v}$
in Eq. (\ref{18}). They are diagonal in the spin variable
$j$. Assuming that all the quantized operators
in terms of $(b,b^*)$ are written in the normal form 
we easily conclude that
\begin{eqnarray}
&(-\mu v_0)_F==\mu\sum_{j}\int\nolimits v_0
b({\bf v},j)^*b({\bf v},j)d\rho({\bf v})\nonumber\\
&(-\mu {\bf v})_F==\mu\sum_{j}\int\nolimits {\bf v}
b({\bf v},j)^*b({\bf v},j)d\rho({\bf v})
\label{39}
\end{eqnarray}

Consider now the operators in Eq. (\ref{26}). Let $A$ be
some of these operators and 
$A({\bf v},j,\epsilon;{\bf v}',j',\epsilon')$ be its
kernel. Since $A$ is diagonal in the spin variable
$j$, it follows from Eqs. (\ref{30}), (\ref{33}) and
(\ref{34}) that the action of $A_F$ on functions with
the carrier in $X_-$ can be written as
\begin{equation}
A_F=\sum_{j}\int\nolimits\int\nolimits
A({\bf v},j,-1;{\bf v}',j,-1)
b({\bf v},j)b({\bf v}',j)^*
d\rho({\bf v})d\rho({\bf v}')
\label{40}
\end{equation}
As noted in Sect. \ref{S4}, the kernel of the
operator $A$ is antisymmetric. By using this fact and
Eq. (\ref{36}), we conclude that Eq. (\ref{40}) can
be rewritten as
\begin{equation}
A_F=\sum_{j}\int\nolimits\int\nolimits
A({\bf v},j,-1;{\bf v}',j,-1)
b({\bf v},j)^*b({\bf v}',j)
d\rho({\bf v})d\rho({\bf v}')
\label{41}
\end{equation}
In other words, the operator $A_F$ has the same form
in terms of $(a,a^*)$ and $(b,b^*)$ operators.

Finally, consider those operators in Eq. (\ref{18})
which contain the $l$th component of the spin
operator ${\bf s}$. Again,   
let $A$ be some of these operators and 
$A({\bf v},j,-1;{\bf v}',j',-1)$ be the part of its
kernel which is of interest for us. It is clear 
from Eq. (\ref{18}), that in that
case $A$ is diagonal in ${\bf v}$, i.e. its kernel
contains $v_0\delta ({\bf v}-{\bf v}')$. Therefore
we can write the kernel in the form
\begin{equation}
A({\bf v},j,-1;{\bf v}',j',-1)=
v_0\delta ({\bf v}-{\bf v}')f({\bf v})
s_{jj'}^l
\label{42}
\end{equation}
where $f({\bf v})$ is a function of ${\bf v}$ and
$s_{jj'}^l$ is the matrix element of $s^l$ for the 
transition between the spin states $j$ and $j'$.
By using Eq.
(\ref{30}) we now obtain that the action of $A_F$ 
on functions
with the carrier in $X_-$ is given by
\begin{equation}
A_F=\sum_{j,j'}\int\nolimits f({\bf v})s_{jj'}^l
a({\bf v},j,-1)^*a({\bf v},j',-1)d\rho({\bf v})
\label{43}
\end{equation}
As follows from Eq. (\ref{33}), in terms of the 
$(b,b^*)$ operators this expression reads
\begin{equation}
A_F=\sum_{j,j'}\int\nolimits f({\bf v})s_{jj'}^l
b({\bf v},-j)b({\bf v},-j')^* \eta (j) \eta (j')^*
d\rho({\bf v})
\label{44}
\end{equation}
Since the trace of any spin operator equals zero,
then by using Eq. (\ref{36}), we can rewrite this expression
as
\begin{equation}
A_F=-\sum_{j,j'}\int\nolimits f({\bf v})s_{jj'}^l
b({\bf v},-j')^*b({\bf v},-j) \eta (j) \eta (j')^*
d\rho({\bf v})
\label{45}
\end{equation}

Consider first the case $l=3$, i.e. $A$ contains the
$z$ component of the spin operator. Since this
component is diagonal in the spin index $j$,
and $j$ is the eigenvalue of the operator $s^3$, it
follows from Eq. ({\ref{34}) that 
\begin{eqnarray}
&A_F=-\sum_{j}\int\nolimits f({\bf v})j
b({\bf v},-j)^*b({\bf v},-j)d\rho({\bf v})=\nonumber\\
&\sum_{j}\int\nolimits f({\bf v})j
b({\bf v},j)^*b({\bf v},j)d\rho({\bf v})
\label{46}
\end{eqnarray}
We conclude that the operators containing the $z$
component of the spin operator have the same form
in terms of $(a,a^*)$ and $(b,b^*)$.

Consider now the operators containing $s^l$
where $l=1$ or $l=2$. 
We choose $\eta(j)$ in the form 
$\eta(j)=(-1)^{(s-j)}$. Then, as follows from
Eq. (\ref{45})
\begin{equation}
A_F=\sum_{j,j'}\int\nolimits f({\bf v})(s)_{jj'}^l
b({\bf v},-j')^*b({\bf v},-j) 
d\rho({\bf v})
\label{47}
\end{equation}
since the operator $s^l$ has nonzero matrix elements 
only for transitions with $j=j'\pm 1$. As follows from
this expression, the operator $A_F$ will have the same
form in terms of $(a,a^*)$ and $(b,b^*)$ if
\begin{equation}
(s)_{j,j'}^l=(s)_{-j',-j}^l
\label{48}
\end{equation}
In the case $s=1/2$ this relation can be easily verified
directly. In the general case it can be proved by using
the properties of $3j$ symbols (see e.g. Ref. \cite{LL}).
Therefore all the operators containing the components of
${\bf s}$ have the same form in terms of $(a,a^*)$ and
$(b,b^*)$.

Our conclusion is as follows. If  
$A=\mu v_0$ or $A=\mu {\bf v}$ then the operator
$A_F$ has the same form in terms of $(a,a^*)$ as
$-A_F$ in terms of $(b,b^*)$. At the same time, the
other operators in Eq. (\ref{18}) have the same form
in terms of $(a,a^*)$ and $(b,b^*)$. 

This result can be reformulated by saying that the 
quantized operators (\ref{18}) can be obtained by
quantizing operators (\ref{19}) with $(b,b^*)$ in
place of $(a,a^*)$. Since the operators (\ref{19})
satisfy the required commutation relations (see the
discussion in Sect. \ref{S4}), we conclude
that for fermions the transformation defined by
Eqs. (\ref{33}) and (\ref{34}) is compatible with
the commutation relations (\ref{2}). 

\section{Discrete symmetries and nonexistence of
neutral elementary particles}
\label{S7}

Let us now discuss the following question. For
definiteness we assumed that $\mu > 0$. Will we
get new UIRs if $\mu < 0$? In Poincare and AdS
theories the choice of the mass sign implies
simultaneously the choice of the energy sign and 
UIRs with the different mass signs are different. 
However, as we have seen in the
preceding sections, in the dS case each UIR contains
the states with both positive and negative 
eigenvalues of the operator $M_{04}$. A well known
result (see e.g. Ref. \cite{Men} for details) is that
UIRs characterized by $\mu$ and $-\mu$ are unitarily
equivalent. For this reason Mensky has proposed the 
following approach for distinguishing
particles from antiparticles: they are described by the
same UIRs but have different space-time interpretation
(see Ref. \cite{Men} for details). In this approach the
UIRs are interpreted in the standard way (see Sect.
\ref{S3}).
 
The fact that the same UIR of the dS group 
contains the states with both the positive and negative 
eigenvalues of the operator $M_{04}$ is the reason of
why the state of the object described by an UIR 
is characterized not only by the velocity ${\bf v}$ 
and the spin projection $j$ but also by a new quantum
number $\epsilon$. In our approach one UIR describes
a particle and its antiparticle simultaneously.
For definiteness we assumed that
$\epsilon =1$ for particles and $\epsilon =-1$ for
antiparticles but such a choice is obviously the
matter of convention. 

As it has been noted in Sect. \ref{S1}, in the
standard theory the fact that a particle and its
antiparticle have equal masses, spins and lifetimes
is a consequence of CPT invariance. Let us discuss
this problem in greater details.

P invariance is described by a unitary transformation, 
which changes the signs of all three dimensional 
momenta. In the standard theory one cannot define T 
invariance 
analogously since in that case the energy sign would 
change. There exist two well known solutions of this
difficulty: the Wigner formulation which involves 
antiunitary operators and the Schwinger
formulation which involves transposed operators \cite{AB}.

The comparison of Eqs. (\ref{17}) and (\ref{18}) shows 
that the operators
$M_{ab}$ in these expressions not containing the 
subscript 4 are the same while those containing 
this subscript have different signs. If the 
coordinates $x^{\nu}$ ($\nu=0,1,2,3$)
are inverted (i.e. one applies the PT transformation)
and no antiunitary or transposed operators are used
then the operators $M_{\nu 4}$ change their
signs while the other operators remain unchanged.
For these reasons one might
think that the operators in Eq. (\ref{18}) are obtained
from ones in Eq. (\ref{17}) by using the PT 
transformation. However, these equations
have been obtained by considering only the elements
of the SO(1,4) group belonging to its unity component, 
and no discrete transformations have been used.
The matter is that the unitary representation operator
corresponding to 
$I$ necessarily changes the sign of the operator $M_{04}$ in 
Eq. (\ref{17}), i.e. it mimics the T transformation. Then 
if ${\bf v}$ is replaced by $-{\bf v}$ we obtain Eq.
(\ref{18}). To overcome the difficulty that the operator
$M_{04}$ in this expression is negative definite in the
Poincare approximation, we relate Eq. (\ref{17}) to 
particles, Eq. (\ref{18}) to
antiparticles and quantize these expressions in a 
proper way. In other words, our analog of the PT 
transformation is accompanied by transition from 
particles to antiparticles, i.e. it is replaced by an 
analog of the CPT transformation. 

As noted by Mensky \cite{Men}, dS invariant theory
could be a basis for new approaches to the CPT 
theorem. We believe that our approach is in the
spirit of Mensky's idea. An analogy between our
approach and the standard CPT transformation is
seen from the following observation. 
In the standard theory the CPT transformation
in Schwinger's formulation transforms the operators 
$b$ to $a^*$ in the transposed form \cite{AB}, and for 
this reason one might think that Eq. (\ref{33}) is the
standard CPT transformation. However, in the standard theory
the operators $a$ and $b$ refer to objects described by
different UIRs while in our approach they refer to the
same object. While in the standard theory the CPT
transformation is a true transformation relating two
sets of physical operators, Eq. (\ref{33})
is not a transformation but {\it a definition} of the
physical $b$ operators in terms of unphysical $a$ operators. 
In particular, $a^*=\eta b$ necessarily
implies $a=\eta^* b^*$ and not $a=\eta b^*$ (i.e.
there is no analogue of antiunitary transformation). 
 
The operators $a({\bf v},j)==a({\bf v},j,1)$ and
$a({\bf v},j)^*==a({\bf v},j,1)^*$ on one hand
and $b({\bf v},j)$ and $b({\bf v},j)^*$ on the
other satisfy the same commutation relations.
As shown in Sect. \ref{S5}, the quantized
representation generators for a particle are obtained
from the operators ({\ref{17}) and $(a,a^*)$. At the
same time, the main result of Sect. \ref{S6} is that
the quantized representation generators for the
corresponding antiparticle are obtained
in the same way but with the operators (\ref{19})
in place of the operators (\ref{17}) and the
operators $(b,b^*)$ in place of the operators
$(a,a^*)$. Since the operators (\ref{17}) and
(\ref{19}) are different (they coincide only in
the limit $R\rightarrow\infty$), we conclude that
different values of $\epsilon$ describe different
sets of representation generators in the quantized
form. Below we discuss this feature in detail since
it has no analogue in the standard theory.

We first show that Eqs. (\ref{17}) and (\ref{19})
are in agreement with the well known results of
General Relativity (GR). For simplicity we consider 
the operators $M_{\nu 4}$ ($\nu=0,1,2,3$) in the
nonrelativistic classical approximation. Denote
${\bf P}={\bf B}/R$, $E=M_{04}/R$, $\mu = mR$, 
${\bf p}=m{\bf v}$ and ${\bf r}=i\partial /\partial {\bf p}$. 
Then as follows from Eq. (\ref{17})
\begin{equation}
{\bf P}= {\bf p}+m{\bf r}/R,\quad 
E = m+{\bf p}^2/2m +{\bf p}{\bf r}/R
\label{49}
\end{equation}
and, as follows from Eqs. (\ref{19}),  
\begin{equation}
{\bf P}= {\bf p}-m{\bf r}/R,\quad 
E = m+{\bf p}^2/2m -{\bf p}{\bf r}/R
\label{50}
\end{equation}
Therefore in the both cases the classical
nonrelativistic Hamiltonian reads
\begin{equation}
E = m + \frac{{\bf P}^2}{2m}-\frac{m{\bf r}^2}{2R^2}
\label{51}
\end{equation}
Note that ${\bf r}$ is canonically conjugated with
${\bf p}$ by construction. In the approximation
when $R$ is large, the last term in the r.h.s. of
Eq. (\ref{51}) is a small correction and ${\bf r}$
is also canonically conjugated with ${\bf P}$. 

The well known result of GR is that if the metric
is stationary and differs slightly from the
Minkowskian one then in the nonrelativistic
approximation the curved space-time can be
effectively described by a gravitational potential
$\varphi({\bf r})=(g_{00}({\bf r})-1)/2c^2$ where 
$g_{00}$ is the time-time 
component of the metric tensor. As follows from
Eq. (\ref{1}), in the approximation when $R$ is
large, the interval squared is given by
\begin{equation}
ds^2=dx_{\nu}dx^{\nu}-(x_{\nu}dx^{\nu}/R)^2
\label{52}
\end{equation} 
We now express $x_0$ in terms of a new variable $t$
as $x_0=t+t^3/6R^2-t{\bf x}^2/2R^2$. Then 
\begin{equation}
ds^2=dt^2(1-{\bf r}^2/R^2)-d{\bf r}^2-
({\bf r}d{\bf r}/R)^2
\label{53}
\end{equation}
Therefore, the metric becomes stationary and
$\varphi({\bf r})=-{\bf r}^2/2R^2$ in agreement with
Eq. (\ref{51}). 

It is well known that in the dS space there exists
antigravity: the force of repulsion between two
particles is proportional to the distance between them.
This easily follows from Eq. (\ref{51}). We now show for 
illustrative purposes how this result can be obtained 
if ${\bf P}$ and $E$ are expressed in terms of ${\bf p}$
and ${\bf r}$ as in Eqs. (\ref{49}) and (\ref{50}).

Consider a system of two free particles described by the
variables ${\bf p}_j$ and ${\bf r}_j$ ($j=1,2$). Define
the standard nonrelativistic variables
\begin{eqnarray}
&&{\bf P}_{12}={\bf p}_1+{\bf p}_2 
\quad {\bf q}_{12}=(m_2{\bf p}_1-m_1{\bf p}_2)/(m_1+m_2)\nonumber\\
&&{\bf R}_{12}=(m_1{\bf r}_1+m_2{\bf r}_2)/(m_1+m_2)\quad 
{\bf r}_{12}={\bf r}_1-{\bf r}_2
\label{54}
\end{eqnarray}
Then if the particles are described by Eq. (\ref{49}),
the two-particle operators ${\bf P}$ and ${\bf E}$ in
the nonrelativistic approximation are given by 
\begin{equation}
{\bf P}= {\bf P}_{12}+M{\bf R}_{12}/R,\quad 
E = M+{\bf P}_{12}^2/2M +{\bf P}_{12}{\bf R}_{12}/R
\label{55}
\end{equation}
where 
\begin{equation}
M = M({\bf q}_{12},{\bf r}_{12})= 
m_1+m_2 +{\bf q}_{12}^2/2m_{12}+{\bf q}_{12}{\bf r}_{12}/R
\label{56}
\end{equation}
and $m_{12}$ is the reduced two-particle mass.
Comparing Eqs. (\ref{49}) and (\ref{55}), we conclude that
$M$ has the meaning of the two-body mass and therefore
$M({\bf q}_{12},{\bf r}_{12})$ is the internal two-body
Hamiltonian. As follows from Eq. (\ref{56}), the classical
equations of motion corresponding to this Hamiltonian
imply that $d^2{\bf r}_{12}/dt^2={\bf r}_{12}/R^2$, i.e.
the internal Hamiltonian (\ref{56}) indeed describes the
dS antigravity.

For a system of two antiparticles the result is
obviously the same since Eq. (\ref{50}) can be formally
obtained from Eq. (\ref{49}) if $R$ is replaced by $-R$.
At the same time, in the case of a particle-antiparticle
system a problem with the separation of external and
internal variables arises. In any case the standard result
can be obtained by using Eq. (\ref{51}).

The above discussion shows that the set of operators 
given by Eq. (\ref{17}) and Eq. (\ref{19}) are 
compatible with GR. At the same time, these sets are
not obviously the same. Recall that the $(a,a^*)$ and
$(b,b^*)$ operators describe annihilation and creation
of particles and antiparticle in the states with a
given velocity ${\bf v}$. If we accept the
main postulate of quantum theory that any selfadjoint
operator represents a measurable physical quantity then
the quantities defined by ${\bf p}$, ${\bf P}$ and $E$
are measurable, at least in principle. Then we conclude
that for particles and antiparticles the operators
${\bf P}$ and $E$ are expressed in terms of ${\bf p}$
differently. With our convention for the choice of the
quantum number $\epsilon$, Eqs. (\ref{49}) and (\ref{50})
for both particles and antiparticles can be written as  
\begin{equation}
{\bf P}= {\bf p}+\epsilon m{\bf r}/R,\quad 
E = m+{\bf p}^2/2m +\epsilon {\bf p}{\bf r}/R
\label{57}
\end{equation}
Note also that in terms of ${\bf v}$ the Lorentz 
group generators in Eqs. 
(\ref{17}) and (\ref{19}) are the same. Therefore
if ${\bf v}$ is expressed in terms of ${\bf P}$ for
particles and antiparticles then these expressions
will become different.

In standard theory the C transformation is defined
as $a\leftrightarrow \eta_C b$ where $\eta_C$ is the
charge parity such that $|\eta_C|^2=1$. Since the
sets of the operators (\ref{17}) and (\ref{19}) are
different, the operators $(M^{(+)}_{ab})_F$ do not
transform into $(M^{(-)}_{ab})_F$ under the C transformation
and {\it vice versa}. Therefore our conclusion is
as follows
\begin{itemize}
\item {\it Statement 2:} Even in the free massive dS 
invariant theory the C invariance is not exact.
\end{itemize}

It is easy to show that the free massive dS theory is 
P invariant and therefore the CP invariance in this 
theory also is not exact.

In the literature (even the very serious one - see e.g.
Okun's book \cite{Okun}) the following question is
sometimes discussed. Suppose that a spaceship from an
extraterrestrial civilization is approaching the Earth,
and the aliens ask us whether the Earth is built of
matter or antimatter. Fortunately, since CP invariance
in weak interactions is not exact, we can explain them 
that the Earth is
built of matter, not antimatter. For example, we can tell
them that the probability to find positrons in the 
$K^0_L$ meson decays is greater than the probability
to find electrons. The above consideration shows that in
the dS invariant theory this could be explained simpler.  

In particle physics a particle is called neutral if it 
indistinguishable from its antiparticle. In particular, 
the C transformation transforms a neutral particle into 
itself. Suppose that the object described by 
an UIR is characterized
by an additive quantum number $q$ such that if $Q$ is the
operator of this number and $\Phi$ is a state such that
$Q\Phi = Q_0\Phi$ then $a^*\Phi$ is a state where
$Qa^*\Phi =(Q_0+q)a^*\Phi$. Then, as follows from Eq.
(\ref{33}), a particle and its antiparticle automatically
have opposite quantum numbers. However, if all the 
additive quantum numbers are equal to zero, we cannot use
this criterion to distinguish a particle from its
antiparticle. Nevertheless, it follows from the above
discussion that in our approach 

\begin{itemize}
\item {\it Statement 3:} Any elementary particle 
cannot be neutral.
\end{itemize}

\section{Discussion}
\label{S8}

In the present paper we have reformulated the 
standard approach
to quantum theory as follows. Instead of requiring that
each elementary particle is described by its own UIR of
the symmetry algebra, we assume Single-Representation
Supposition (see Sect. \ref{S1}) that one UIR should 
describe a particle and its antiparticle simultaneously.
In that case, among the Poincare, AdS and dS algebras,
only the latter can be a candidate for constructing
elementary particle theory. 

We show in Sect. \ref{S3} that UIRs of the dS algebra 
cannot be interpreted in the standard way since such an
interpretation is physically meaningless even in
the Poincare approximation. Although our approach 
considerably differs from LQFT in curved
spacetime, this conclusion is in agreement with
that in Ref. \cite{Narlikar} and references 
therein. 

The important ingredient of our construction is that
nonphysical states are associated with antiparticles.
In Ref. \cite{Susskind} problems with the dS theories
have been discussed in the framework of the 
thermofield theory which was developed 
many years ago in many-body theory (see 
\cite{Susskind} for references). In this theory 
there also exist physical and nonphysical states
but our interpretation is essentially different. 
Nevertheless, it is
interesting to note that similar ideas 
can work in approaches which considerably
differ each other.  

Although Single-Representation Supposition seems
natural, it is a supposition. Therefore the
problem arises whether it can be substantiated.
As noted in Sect. \ref{S1}, the present investigation
has been inspired by our results in quantum theory
over a Galois field (GFQT) \cite{lev2}. Here 
no analog of Single-Representation Supposition is
needed since any IR of the symmetry algebra 
automatically describes a particle
and its antiparticle simultaneously.

The main idea of our approach is
extremely transparent: {\it if there exists any
criterion for separating physical and nonphysical 
states then only fermions can be elementary}.
Indeed, the transition from nonphysical to 
physical states involves replacement of annihilation
operators by creation ones and {\it vice versa}.
Therefore if $a =\eta^* b^*$ then necessarily 
$a^* = \eta b$, $\{a,a^*\}=\eta \eta^* \{b,b^*\}$ 
and $[a,a^*]=-\eta \eta^*[b,b^*]$. It is easy to 
satisfy the condition $\eta\eta^* =1$ but in the 
field of complex numbers it is impossible to satisfy the
condition $\eta\eta^* =-1$.  

The criterion used in the present work is that the
energy should be positive definite in the Poincare 
approximation.
Let us discuss this question in greater details.

The dS group contains transformations which 
transform positive energy states to negative energy
ones and {\it vice versa}. This does not
necessarily represent a problem. As we have already
mentioned, the standard treatment of UIRs does not
apply in the dS case. Note that our results are
based only on the properties of representation
generators while the dS space has not been used at 
all (in Sect. \ref{S7} we mentioned the dS space 
only for illustrative purposes to show that the results
are compatible with GR when $R$ is large). As noted in 
Sect. \ref{S1}, the classical space-time cannot be
fundamental in quantum theory (see e.g. Ref. \cite{gravity}
for a detailed discussion; this conclusion is also in 
the spirit of
Heisenberg's S-matrix program). In any case, in the 
approximation when $R$ is large, one UIR of the dS algebra 
asymptotically splits into well known disjoint UIRs of 
the Poincare algebra with positive and negative energies,
respectively. Therefore, Single-Representation 
Supposition in this approximation is fully consistent. 

In addition to Statement 1 (see Sect. \ref{S5}) that 
only fermions can be elementary, it also follows from our 
consideration (see Statement 2 and Statement 3 in Sect. 
\ref{S7}) that 
\begin{itemize}
\item Even in the free massive dS 
invariant theory the C invariance is not exact.
\item Any elementary particle cannot be neutral.
\end{itemize}
Statement 1 and Statement 3 give a natural explanation
of the fact that neutral fermions have not been
observed and all observed neutral states are bosons.

The famous Pauli spin-statistics theorem 
\cite{Pauli} in LQFT states
that fermions necessarily have a half-integer spin
while bosons - an integer spin. After the original 
Pauli proof, many authors investigated more general 
approaches to the spin-statistics theorem (see e.g. Ref. 
\cite{Kuckert} and references therein). 
Since in our approach only fermions can be elementary,
the problem arises whether it is possible to prove
that their spin is necessarily half-integer. 
In Ref. \cite{lev2} we have considered Galois
field analogs of IR of the so(2,3) algebra.
It has been shown that in the GFQT, as a consequence
of simple arithmetic considerations, the vacuum
condition is consistent only for particles with
a half-integer spin. At the same time, we did not
succeed in proving that only fermions can be
elementary. The matter is that in Galois fields
the relation $\eta \eta^* =-1$ is not impossible.

As noted in Sect. \ref{S2}, the explicit description
of the representation space and operators depends on
the choice of representatives in a certain coset space.
The choice adopted in the present paper is convenient
in the Poincare approximation but when $R$ is not 
asymptotically large there exist more
natural choices (see e.g. Refs. 
\cite{Men,lev1,lev3,gravity}). Recall that
in our approach one UIR describes an object, and,
proceeding from our experience, we wish to
separate the states of that object into those 
related to either a particle or its antiparticle.
In other words, if $H$ is the representation space 
then we wish to find subspaces $H_+$ and $H_-$ such that
$H$ is a direct sum of $H_+$ and $H_-$, $H_+$ represents
all possible states for a particle and $H_-$ - all 
possible states for its antiparticle. The 
results of Refs. \cite{lev2,spin} show that in general, 
whatever separation criterion is used, representation 
generators have nonzero matrix elements for transitions 
between $H_+$ and $H_-$.
Is this an indication that the 
very notion of particles and antiparticles has exact
meaning only in the Poincare approximation? In 
particular, does 
the conservation of electric charge take place only in 
the Poincare approximation? These problems deserve 
further study.

The possibility that only fermions
can be elementary is very appealing from the
aesthetic point of view. Indeed, what was the reason for
nature to create elementary fermions and bosons if the 
latter can be built of the former? A well known
historical analogy is that before the discovery of
the Dirac equation, it was believed that nothing could
be simpler than the Klein-Gordon equation for spinless
particles. However, it has turned out that the spin 1/2
particles are simpler since the covariant equation
for them is of the first order, not the second one as the
Klein-Gordon equation. A very interesting possibility
(which has been probably considered first by
Heisenberg) is that only spin 1/2 particles could be
elementary.

{\it Acknowledgements: } I have greatly benefited from 
discussions with many physicists and mathematicians, and 
it is difficult to mention all of them. 
A collaboration with L.A. Kondratyuk and discussions with 
S.N. Sokolov were very
important for my understanding of basics of quantum theory.
They explained that the theory should not necessarily be 
based on a local Lagrangian, and symmetry on quantum level 
means that proper commutation relations are satisfied. 
E.G. Mirmovich has proposed an idea that only angular momenta 
are fundamental physical quantities \cite{Mirmovich}. This 
has encouraged me to study de Sitter invariant 
theories. At that stage the excellent book by Mensky
\cite{Men} was very helpful. I am also grateful to F. Coester, 
V. Dobrev, M. Fuda, B. Hikin, B. Kopeliovich, L. Koyrakh, 
M. Olshanetsky, E.Pace, W. Polyzou and G. Salme for 
numerous discussions.

\begin{sloppypar}
The manuscript of this paper has been refereed by three anonymous 
referees: First Referee, Second Referee and Adjudicator. I am very 
grateful to First Referee and Adjudicator for the support
of this work and important critical remarks. Second Referee's
remarks concerning definition of energy in the dS 
theory were very helpful for the discussion 
in Sect. \ref{S3}.
\end{sloppypar}

\end{document}